\documentclass[12pt]{article}
\input epsf
\usepackage{amsmath}
\usepackage{amssymb}
\usepackage[final]{graphicx}
\usepackage{cite}

\newcounter{eqlett}

    \def\be{\begin{eqnarray}}
    \def\ee{\end{eqnarray}}
    \def\no{\nonumber}

    \def\suml{\sum\limits}
    \def\prodl{\prod\limits}
    \def\intl{\int\limits}

    \def\bn{\begin{enumerate}}
    \def\en{\end{enumerate}}
    \def\bi{\begin{itemize}}
    \def\ei{\end{itemize}}

    \def\({\left(\!}
    \def\){\!\right)}
    \def\<{\left\langle\,}
    \def\>{\, \right\rangle}
    \def\[{\left[}
    \def\]{\right]}

    \def\tilde{\widetilde}

    \def\a{\alpha}
    \def\b{\beta}
    \def\d{\delta}
    \def\g{\gamma}
    \def\G{\Gamma}
    \def\e{\epsilon}

    \def\l{\lambda}

    \def\p{\phi}

    \def\CN{{\cal N}}
    \def\CO{{\cal O}}
    
    \def\CQ{{\cal Q}}

    \def\p{\partial}

    \def\MC{{\mathbb{C}}}

    \def\MR{{\mathbb{R}}}

    \def\MZ{{\mathbb{Z}}}

    \def\sign{{\rm{sign}}\,}

    \def\Tr{{\rm Tr}\,}

    \def\sl{\textit{sl}}

    \def\xpm{ x^{\pm} }
    \def\xp{ x^{+} }
    \def\xm{ x^{-} }
    \def\ypm{ y^{\pm}}
    \def\yp{ y^{+}}
    \def\ym{ y^{-}}

\begin{document}

\thispagestyle{empty}
\begin{flushright}
{ }
{IPhT-T08/xxx}\\
\end{flushright}
\vspace{1cm}
\setcounter{footnote}{0}
\begin{center}
{\Large{
\bf
The 2-loop generalized scaling function from the BES/FRS equation
}}

\vspace{20mm} {Dmytro Volin} \\[7mm] {\it Institut de Physique Th\'eorique, CNRS-URA
2306
\\
C.E.A.-Saclay \\
F-91191 Gif-sur-Yvette, France}
\\[7mm] {\it Bogolyubov
Institute for Theoretical Physics, 14b Metrolohichna Str.  \\
Kyiv, 03143 Ukraine}.
 \end{center}

\vskip 9mm

\noindent{We formulate the BES/FRS equation as a functional equation in the rapidity space and perform its strong coupling expansion in the limit when $\ell=j/4g$ is kept finite. We obtain a result which is consistent with the previous calculations at tree level and one loop and which differs from the two-loop calculation in 0805.4615 by a term singular at $\ell=0$.}
\vfill

\newpage

\section{Introduction and main results}
An important test of the AdS/CFT correspondence is based on the comparison of the anomalous dimension of the Wilson twist $L$ operator \cite{Belitsky:2006en} and the energy of the folded string spinning  on AdS$_3\times$S$^1$. The general twist $L$ operator has the following form:
\begin{eqnarray}\label{twistL}
\Tr(D_+^{n_1}Z
D_+^{n_2}Z\ldots D_+^{n_L}Z),\;\;\;\;&& {n_1\!\!+\!n_2\!\!+\!\ldots\!\!+\!n_L\!\!=\!M},
\end{eqnarray}
where $D_+$ is the light cone covariant derivative and
$Z={\mit\Phi}_1+i {\mit\Phi}_2$ is a complex scalar field.

In the spin chain picture the fields $Z$ correspond to the nodes of the $\sl(2)$ spin chain and the covariant derivatives $D$ to the excitations (magnons).
The anomalous dimension of the operator (\ref{twistL}) plays the role of the energy of the corresponding spin chain state. The energy of the ground state $\g[g,L,M]$ is a function only of the gauge coupling constant $g=\frac{\sqrt{\l}}{4\pi}$, the length $L$ and the number of magnons $M$. This function admits a remarkable logarithmic scaling when $M\to\infty$ \cite{Belitsky:2006en}:
\be\label{logm}
  \gamma=f[g,\ldots]\log M+\ldots,
\ee
which is valid only for certain scaling behavior of $L$: at weak coupling $L\lesssim \log M$ and at strong coupling for $L\lesssim g \log M$.

The important case with $L/\log M\to 0$ was initially considered \cite{Belitsky:2006en,Eden:2006rx}. In this case the prefactor $f[g]$ in (\ref{logm}) depends only on the coupling constant and is equal to twice the cusp anomalous dimension $\G_{cusp}[g]$. The cusp anomalous dimension is an object which can be defined for any four-dimensional gauge theory. For $\CN=4$ SYM it was computed up to four loops \cite{Bern:2006ew}.

In the limit $M\to\infty$, $L/\log M\to 0$ the Bethe Ansatz equations which describe the $sl(2)$ spin chain can be reduced to the linear integral equation, known as the BES equation \cite{Eden:2006rx,Beisert:2006ez}. This equation allows to find the cusp anomalous dimension at any value of the coupling constant. The condition that the weak coupling expansion of the solution matches with the 4-loop calculations in the gauge theory side  was an important ingredient for establishing the complete form of the Bethe equations. Using the BES equation, the strong coupling expansion of the cusp anomalous dimension was performed numerically \cite{Benna:2006nd} and analytically at the leading \cite{Kotikov:2006ts,Alday:2007qf,Kostov:2008ax,Beccaria:2007tk} order. In \cite{Basso:2007wd,Kostov:2008ax} a recursive procedure for analytical expansion to any desired order was given. The obtained results reproduced the string theory calculations at tree level \cite{Gubser:2002tv,Frolov:2002av}, one \cite{Frolov:2006qe} and two \cite{Roiban:2007dq} loops. The first three orders of the strong coupling expansion are given by
\be
  2\G_{cusp}=\frac 1\e-\frac {3\log[2]}{\pi}-\e\frac{K}{\pi^2}+\CO(\e^2),\;\;\;\;\e=\frac 1{4g}\;.
\ee
The cusp anomalous dimension is an example of a function which smoothly interpolates between weak and strong coupling regimes. Freyhult, Rej and Staudacher \cite{Freyhult:2007pz} proposed a more general function which might had this property. They considered the limit
\be\label{introFRS}
  M,L\to \infty,\hspace{1.5EM}j=\frac {L}{\log M}\hspace{0.5EM}\hspace{0.5EM}\textrm{fixed}\;
\ee
and showed that in this case the logarithmic scaling (\ref{logm}) exists at all orders of the perturbation theory . The limit (\ref{introFRS}) was initially introduced in \cite{Belitsky:2006en} and the logarithmic scaling at $g=0$ and arbitrary $j$ was observed in \cite{Alday:2007mf}. In the limit (\ref{introFRS}) the function $f^{FRS}[g,j]$ depends on two parameters and is called the generalized scaling function. It can be extracted from the solution of the linear integral equation (the BES/FRS equation), introduced in \cite{Freyhult:2007pz}.

On the string theory side the logarithmic scaling is in particular realized in the following limiting procedure, which is taken in two steps:
\be\label{stringlimit}
1)\;\;&& g\!\sim\! L\!\sim\! M\to\infty,\no\\
2)\;\;&& M/g\gg L/g\gg 1,\;\; \ell=\frac{L}{4g\log [M/g]} \;\;\;\textrm{finite}.
 \ee
The prefactor of $\log M$, which we will denote as $f^{string}$, is given by the following strong coupling expansion
\be\label{eq:scs}
 f^{string}[\e,\ell]=\frac 1{\e}(f_0^{string}[\ell]+\e f_1^{string}[\ell]+\e^2 f_2^{string}[\ell]+\ldots).
\ee
Comparing the limits (\ref{introFRS}) and (\ref{stringlimit}), Freyhult, Rej and Staudacher raised the question whether the strong coupling expansion of $f^{FRS}[g,j]$ in the limit $g,j\to\infty$ and $\ell=j/4g$ fixed coincides with $f^{string}[\e,\ell]$.

From the string theory perturbative calculations the tree \cite{Gubser:2002tv,Frolov:2002av} and the one loop \cite{Frolov:2006qe} results for the expansion (\ref{eq:scs}) were obtained for any value of $\ell$. At two loops only the first two orders of the small $\ell$ expansion were found \cite{Roiban:2007ju}:
\be\label{Tseytlin}
  f^{string}_2(\ell)\!\!&=&\!\!-\frac{K}{\pi^2}+\frac{\ell^2}{\pi^2}(q^{string}_{02}-6\log\ell+8\log^2\ell)+{\cal{O}}(\ell
^4)\,,
\ee
where $q^{string}_{02}=2K-\frac 32\log 2+\frac {7}4$.

The limit (\ref{stringlimit}) was also used for the calculations from the asymptotic Bethe Ansatz. In this limit the expansion (\ref{eq:scs}) for arbitrary value of $\ell$ was found at tree and one loop \cite{Casteill:2007ct,Belitsky:2007kf} and then at two loop order \cite{Gromov:2008en}.
While tree and one loop calculations coincide with the string theory predictions, the small $\ell$ expansion of the two-loop result \cite{Gromov:2008en} is different and is given by\footnote{We use normalization which is different from the one used in \cite{Gromov:2008en}.}:
\be\label{Gromov}
  f^{BA}_2(\ell)\!\!&=&\!\!-\frac{K}{\pi^2}+\frac{\ell^2}{\pi^2}(q^{BA}_{02}-6\log\ell+8\log^2\ell)+{\cal{O}}(\ell
^4)
\ee
with $q^{BA}_{02}=-\frac 32\log 2+\frac {11}4$.

A different limit, particularly  interesting from the string theory perspective, was proposed by Alday and Maldacena [16]. In this limit $g\to\infty$ and $j$ is exponentially small with respect to $g$:
\be\label{aldaymaldacena}
  j \sim m\sim g^{1/4}e^{-\pi g}.
\ee
In this limit only massless excitations on $S^5$ are important. Therefore, the theory should be described by the O(6) sigma model. The parameter $m$ is identified with the mass gap of the O(6) sigma model. In the limit (\ref{aldaymaldacena}) the difference between the generalized scaling function and twice the cusp anomalous dimension can be expanded in the powers of $j$ and is given by the following expression
\be
  f^{FRS}[g,j]-2\G_{cusp}[g]=-j+j^2 E[j/m]+\CO(j^4)\ldots,
\ee
where the term $j^2 E[j/m]$ is identified with the energy density of the ground state of the $O(6)$ sigma model. The corrections of the order $\CO(j^4)$ cannot be obtained from the $O(6)$ sigma-model.

Basso and Korchemsky \cite{Basso:2008tx} applied the Alday-Maldacena limit to the BES/FRS equation and derived the thermodynamic Bethe Ansatz of the O(6) sigma model at zero temperature. Therefore, the strong coupling expansion of the generalized scaling function in this limit (\ref{aldaymaldacena}) should reproduce the string theory predictions.

The expansion of $E[j/m]$ at small $j/m$ was computed in \cite{Basso:2008tx,Fioravanti:2008ak,Fioravanti:2008bh}. The expansion at large of $j/m$ was done in \cite{Bajnok:2008it} resulting in the following expression for the generalized scaling function:
\be\label{Bajnok}
  &&f^{FRS}[g,j]\!-\!2\G_{cusp}[g]=\no\\&&=-j+\e^2 j^2\!\(\frac 1{2\e}+\frac 1\pi\!\(\frac 32-2\log\ell\)+\frac \e{\pi^2}\(\ q^{BA}_{02}-6\log\ell+8\log\ell^2\ \)\!\)\!\!+\CO(j^4)
\ee
with $q^{BA}_{02}=-\frac 32\log 2+\frac {11}4$.

We see that the results (\ref{Gromov}) and (\ref{Bajnok}) coincide, although they were obtained in the different orders of limits.

The main goal of the current paper is to perform the strong coupling expansion of the generalized scaling function $f^{FRS}[g,j]$ with $\ell=j/4g$ fixed. The order of limits that we use is different from what was used for the calculation of (\ref{Tseytlin}),(\ref{Gromov}) and from what was used for the calculation of (\ref{Bajnok}). At tree level and one loop we obtain the result which coincides with $f_0^{string}[\ell]$ and $f_1^{string}[\ell]$ (and therefore with the Bethe Ansatz calculations in the limit (\ref{stringlimit})). At the two-loop order we obtain the answer for arbitrary $\ell$ which can be written in terms of $f_2^{BA}[\ell]$:
 \be\label{resultintro}
  f_2^{FRS}[\ell]=f_2^{BA}[\ell]+\frac 1{\sqrt{1+\ell^2}}\(\frac 1{24}\frac 1{\ell^6}+\frac 1{12}\frac 1{\ell^4}\).
 \ee
It is interesting to consider the large $\ell$ expansion of $f_2[\ell]$. The reason for this is the following. As we can conclude from \cite{Frolov:2006qe}, at the first few orders the large $j$ expansion of the scaling function $f[g,j]$ should have the BMN-like properties. This means the following. The large $j$ expansion has the form
\be\label{BMNlike}
  f[g,j]=\suml_{n\geq 1}\frac{g^{2n}}{j^{2n}}\suml_{m\geq 0}\frac{c_{nm}[g]}{j^{m+1}}
\ee
 and the coefficients $c_{10},c_{11},c_{12},c_{20},c_{21}$ do not depend on the coupling constant. All these coefficients except the $c_{12}$ can be found from tree \cite{Gubser:2002tv,Frolov:2002av} and one loop \cite{Frolov:2006qe} calculations on the string side. The prediction for them was confirmed by numerical computation at weak coupling \cite{Beccaria:2008nf}. In the current paper we derive the coefficient $c_{12}$ from $f_2^{FRS}[\ell]$. We find that
\be\label{c12}
f_2^{FRS}[\ell]=\frac 13\frac 1{\ell^3}+\CO(\ell^{-5})\;\;\textrm{and}\;\;c_{12}=\frac {16}3.
\ee
We verify this prediction by the numerical computations at weak coupling (see Appendix \ref{sec:largej}). Note that both $f_2^{FRS}[\ell]$ and $f_2^{BA}[\ell]$ give the same prediction for $c_{12}$.

The article is organized as follows. In Sec.~2 we derive the BES/FRS equation from the Baxter-like equation. In Sec.~3 we analyze the analytical properties of the resolvents. In Sec.~4 we formulate the perturbative solution at strong coupling and perform explicit calculations for tree level, one and two loops. For the two-loop order we use the results of Sec.~5, in which we analyze the behavior of the solution near the branch points. Finally, in Sec.~6 we summarize the obtained results.

\section{BES/FRS equation}
We start from the Bethe Ansatz equations for the $\sl(2)$ sector \cite{Beisert:2005fw,Beisert:2006qh,Beisert:2006ez,Arutyunov:2004vx}:
\be\label{sl2bethe}
  \(\frac{x_k^+}{x_k^{-}}\)^L=\prodl_{\substack{j=1 \\j\neq k}}^{M}\frac{u_k-u_j-2i\e}{u_k-u_j+2i\e}\(\frac{1-\frac{1}{x_k^+\xm_j}}{1-\frac{1}{x_k^-\xp_j}}\)^2e^{2i\theta[u_k,u_j]}\;.
\ee
Here $\e=\frac 1{4g}$ and the normalization of rapidities is suited for the strong coupling expansion.

The variable $x$ is the inverse Jukowsky map of $u$:
\be
  u=\frac 12\(x+\frac 1x\),\;\;\;\; x[u]=u\(1+\sqrt{1-\frac 1{u^2}}\),\;\;\;\;\xpm[u]=x[u\pm i\e]\;.
\ee
The branch of the square root is chosen in a way that $|x|>1$.

We assume $M$ to be even and enumerate the Bethe roots in a way that $u_k>u_l$ for $k>l$.

The dressing phase $\theta[u,v]$ can be represented in the form
\be\label{dressingsemisym}
  \theta[u,v]=\frac 12\left( \chi[\xp,\ym]+\chi[\xm,\yp]+\chi[\xm,-\ym]+\chi[\xp,-\yp]\right).
\ee
We used the notation $\ypm=x[v\pm i\e]$. We will also use $y=x[v]$ below.

The function $\chi[x,y]$ is analytic for $|x|>1$ and $|y|>1$ and respects the following parity properties:
\be
  \chi[x,y]=-\chi[y,x]=-\chi[x,-y].
\ee

Since we consider the ground state which is symmetric, the Bethe equations will not change if we replace the dressing phase with the function $\chi[\xp,\ym]+\chi[\xm,\yp]$.

In the following we will use the Baxter-like equation
\be\label{baxtercomplete}
  Q[u]\,T[u]=W[u+i\e]Q[u+2i\e]+W[u-i\e]Q[u-2i\e]
  \ee
introduced in \cite{Belitsky:2007kf}. Here
\be
  Q[u]&=&\prodl_{k=-\frac M2}^{\frac M2}(u-u_k),\no\\
  W[u\pm i\e]&=&(\xpm)^L\prodl_{k=-\frac M2}^{\frac M2}\(1-\frac 1{\xpm{x^\mp_k}}\)^{- 2}e^{\mp 2i\chi[\xpm,x^\mp_k]}.
\ee
The equation (\ref{baxtercomplete}) should be understood as the definition of $T[u]$. The requirement of analyticity of $T[u]$ on the real axis is equivalent to the requirement for $u_k$ to satisfy the Bethe equations (\ref{sl2bethe}) with the dressing phase $\theta[u,v]$ replaced by $\chi[\xp,\ym]+\chi[\xm,\yp]$. The equation (\ref{baxtercomplete}) implies that $T[u]$ has $L$ zeros which are usually called holes. One can show that both the Bethe roots $u_k$ and the holes are real.

The equation (\ref{baxtercomplete}) resembles the Baxter equation for the eigenvalues of the transfer matrix. However, there is no known transfer matrix or equivalent object with eigenvalues given by zeros of $T[u]$.

To proceed, we introduce the resolvents
\be\label{resolventsdefinition}
  R_m[u]&=&\frac 1{\log M}\frac{d}{du}\log Q[u],\no\\
  R_h[u]&=&\frac 1{\log M}\frac{d}{du}\log T[u]
\ee

The distribution of the Bethe roots (magnons) is given by the density function $\rho_m[u]$. It is supported on the two intervals $[-a_{ext},-a]\cup[a,a_{ext}]$ with $a\sim 1$ and $a_{ext}\sim S$. The density $\rho_m[u]$ is related to the resolvent $R_m$:
\be
  \rho_m[u]=-\frac 1{2\pi i}\(\ R_m[u+i0]-R_m[u-i0]\ \).
\ee
As we argue in the appendix \ref{appendixa}, the density is finite at $u=\pm a$ and the resolvent $R_m$ has a \textit{logarithmic} type singularity at these points. In other words, the typical distance $d$ between the roots near $u=\pm a$ approaches zero as $L^{-1}$ in the considered limit. Therefore, $d$ is much smaller than the shift $\e$ in the equation (\ref{baxtercomplete}) for any finite value of $\e$. This implies the fact that $W[u+i\e]Q[u+2i\e]\gg W[u-i\e]Q[u-2i\e]$ for $\Im[u]>0$ and $W[u+i\e]Q[u+2i\e]\ll W[u-i\e]Q[u-2i\e]$ for $\Im[u]<0$.

It is instructive to compare the limit (\ref{introFRS}) with another well-studied limit $L\sim M\to~\infty$. In the latter case $\e\ll d$ and at the distances from the real axis of order $\e$, which are finite, the two terms of the r.h.s of (\ref{baxtercomplete}) have the same magnitude.

In the limit (\ref{introFRS}) all the holes except two are supported on the interval $[-a,a]$. Their distribution is given by the function $\rho_h$ given by the discontinuity of $R_h$ on $[-a,a]$:
\be
  \rho_h[u]=-\frac 1{2\pi i}\(\ R_h[u+i0]-R_h[u-i0]\ \).
\ee
There are two holes which are situated outside the interval $[-a_{ext},a_{ext}]$. The position of these two external gives us the information about the asymptotic behavior of $R_m$ for large absolute values of $u$. One can show that at the scales $u\sim\log M$ the resolvent $R_m$ is constant in the leading order of the limit (\ref{introFRS}) \cite{Eden:2006rx}. In our normalization this constant equals to $-\frac {i}{\e}$. At scales larger than $u\sim\log M$ the roots do not contribute to the leading $\log M$ order of the energy. Therefore, we will consider the scale $u\sim\log M$  as infinity.

The asymptotic behavior of the resolvents is given by
\be\label{escaling}
  R_m&\rightarrow& \mp \frac {i}{\e}+\frac{\beta}u\,,\, u\rightarrow \infty\pm i0;\no\\
  R_h&\rightarrow& \frac{j}{u}\,,\,u\to\infty.
\ee
The generalized scaling function can be found by:
\be\label{f2cj}
  f=-2\beta-j\;.
\ee

In the following we consider $\Im[u]>0$. We can neglect the second term in the r.h.s of (\ref{baxtercomplete}). Taking the logarithmic derivative of
(\ref{baxtercomplete}), we obtain the equation
\be\label{prefrs}
  (1-D^2)R_m+R_h-\frac{d}{du}\frac{\log W[u+i\e]}{\log M}=0,
\ee
where $D$ is the shift operator
\be
  D=e^{i\e\p_u}.
\ee
Using the resolvent $R_m$ we can rewrite the sum over the Bethe roots as the contour integral. In particular, the term in (\ref{prefrs}) containing $\log W[u+i\e]$ can be rewritten as
\begin{small}
\be\label{holononbes}
  -\frac{d}{du}\frac{\log W[u+i\e]}{\log M} = -2 D\intl_{\MR-i0} \frac {dv}{2\pi i} \p_u\(\log\[1-\frac 1{x y}\]+i\chi[x,y]\)DR_m-D\frac{L}{\log M}\frac 1{x}\frac{dx}{du}\;.
\ee
\end{small}
We used the fact that $W[u+i\e]$ as the function of $u_k$ is analytic in the lower half plane.

Further simplification can be achieved by performing the contour deformation, explained in Sec.~3.2 of \cite{Kostov:2008ax}:
\be\label{contourtrick}
  -\!\!\!\!\!\intl_{-\infty-i0}^{\infty-i0}\!\!\!\!\!\frac {dv}{2\pi i}\p_u\log\[1-\frac 1{xy}\]DR_m[v]=\!\!\!\!\intl_{-1+i0}^{1+i0}\!\!\!\!\!\frac{dv}{2\pi i}\frac{y-\frac 1y}{x-\frac 1x}\frac 1{v-u}DR_m[v]=\frac{K_-+K_+}{2}DR_m,
\ee
where $K_\pm$ - the kernels introduced in \cite{Kostov:2008ax}:
\be
  (K_\pm F)[u]\equiv\intl_{-1+i0}^{1+i0}\frac{dv}{2\pi i}\frac{y-\frac 1y}{x-\frac 1x}\frac 1{v-u}\(F[v+i0]\pm F[-v+i0]\).
\ee
We can make the same contour deforming trick with $\p_u \chi$. The conjectured BES/BHL dressing phase \cite{Beisert:2006ez,Beisert:2006ib} is such that\footnote{The expression $\frac 1{1-D^2}$ should be understood as a series over positive powers in D.}
\be\label{predressing}
    \intl_{-\infty-i0}^{\infty-i0}i\p_u\chi[x,y] DR_m[v]\frac {dv}{2\pi i}=-K_-\frac {D^2}{1-D^{2}}K_+DR_m.
\ee
Using (\ref{holononbes}), (\ref{contourtrick}) and (\ref{predressing}), we get from (\ref{holononbes}) the functional version of the BES/FRS equation valid in the upper half plane:
\be\label{holobes}
(1-D^2)R_m+R_h=-D\(K_-+K_++2K_-D\frac 1{1-D^2}DK_+\)D R_m + \frac 1{\e}D \frac {\ell}{x}\frac {dx}{du}
\ee
with $\ell=\e j$.
This equation can be also derived from the original BES/FRS equation, formulated in the Fourier space \cite{Freyhult:2007pz}, by means of an inverse half-Fourier transform \cite{unpublished}. The neglecting the nonlinear term in the derivation of the BES/FRS equation from the nonlinear integral equation \cite{Bombardelli:2008ah} is equivalent to the neglecting one of the terms in the r.h.s of the equation (\ref{baxtercomplete}).

\section{Analytic properties of the resolvents and possible regimes at strong coupling}\label{sec:analytical}
\begin{figure}\label{fig:resolvents}\centering
\includegraphics[width=5cm,height=5cm]{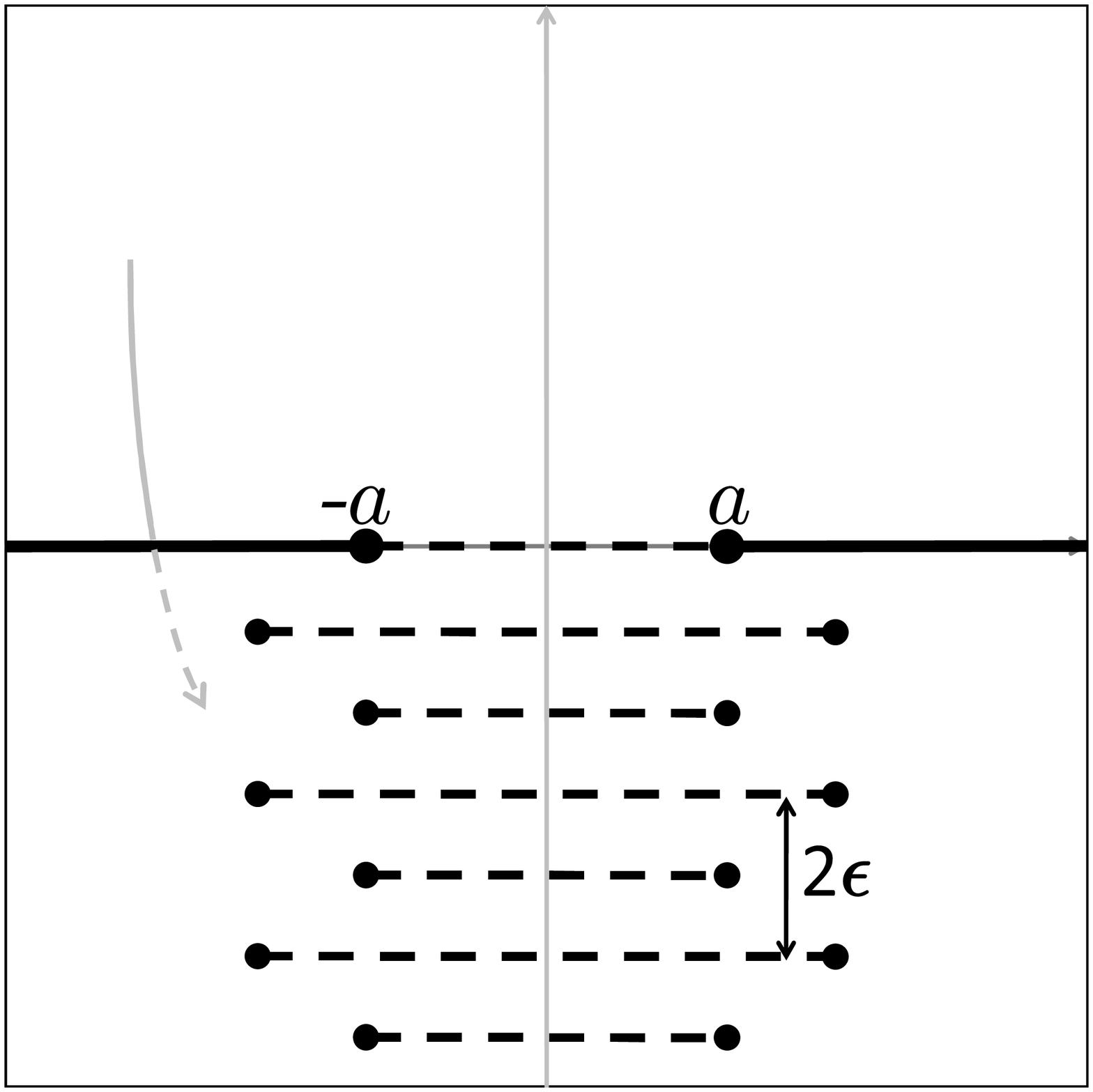}\qquad
\includegraphics[width=5cm]{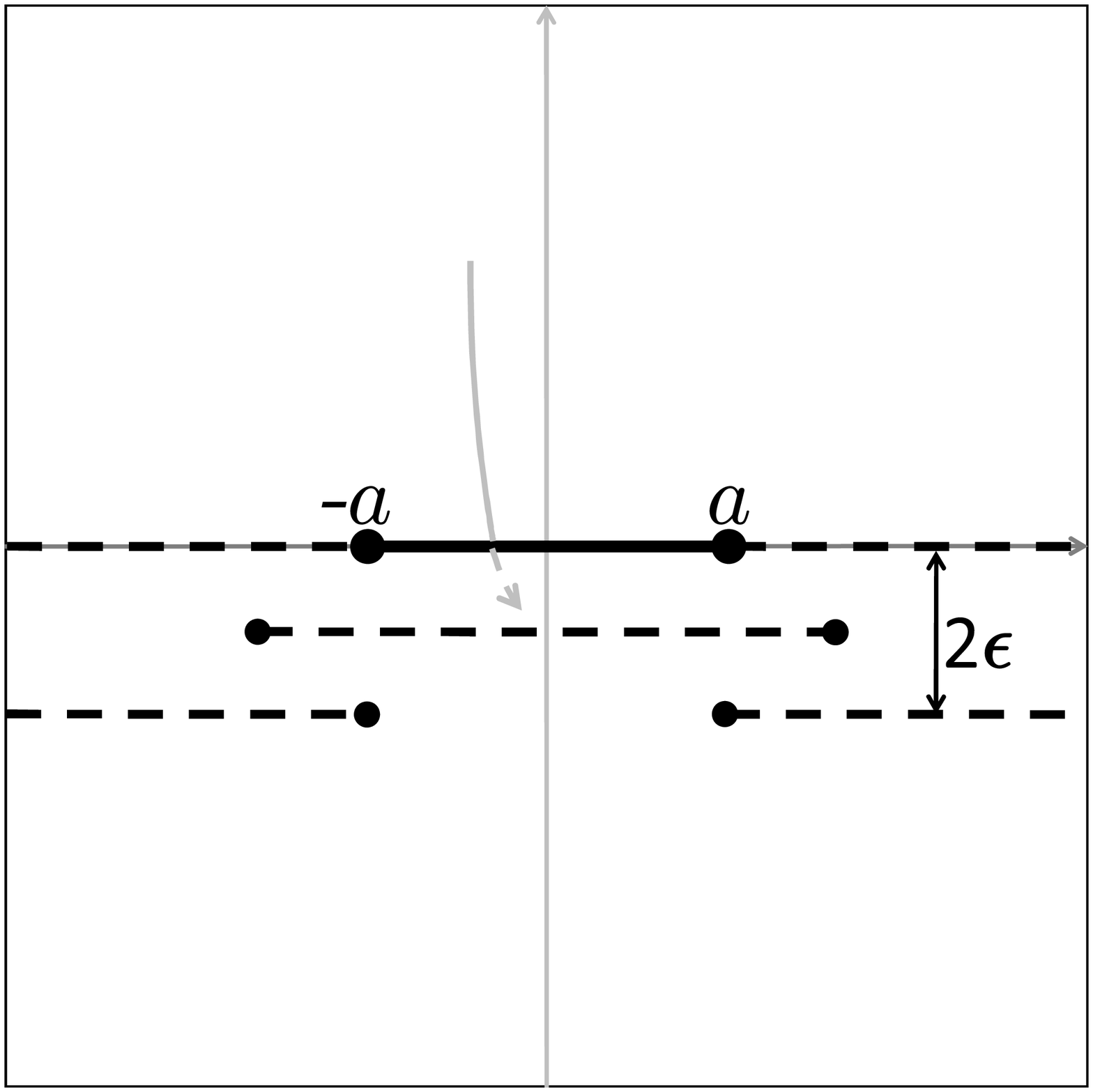}
\vskip 4mm \caption{\small Left: Analytical structure of $R_m[u]$.
Right: Analytical structure of $R_h[u]$. Solid line - the only cut on the physical sheet which corresponds to the roots/holes condensation. Dashed lines correspond to the cuts which appear if we analytically continue the resolvents from the upper half plane by the path shown by arrow.}
\label{fig:cutsrR}
\end{figure}

The analytical structure of the resolvents can be deduced from (\ref{holobes}) and is shown in Fig.~\ref{fig:resolvents}.
There are two types of the branch points of the resolvents. The first type of branch points has "kinematic" origin - these are branch points of the inverse Jukowsky map $x[u]$ which is singular at $u=\pm 1$. As we can deduce from the equation (\ref{holobes}) (and its conjugate which is valid in the lower half plane), the "kinematic" branch points are located at $u=\pm 1+(2\MZ+1)i\e$. The branch points of the second type are located at $u=\pm a+ 2\MZ i\e$, where $a[g,j]$ is the end of the root distribution. The only branch points on the physical sheet are $u=\pm a$. All the others appear after the analytical continuation through the cut to the nonphysical sheets.

In the following we make an assumption that the generalized scaling function is bianalytic function of $g$ and $j$ everywhere except the values of $g$ and $j$ for which the branch points of different type collide. For real $g$ and $j$ the collision is impossible except $g=\infty$. However, if we take into account complex values of these parameters, the collision is possible and it determines the radius of convergence for the Taylor expansion of $f[g,j]$ around some analytical point. For example, for $j=0$ we have $a=0$. The kinematic branch point touches the origin when $g=i(1/4+\MZ/2)$. Therefore, the radius of convergence of the weak coupling expansion of $2\G_{cusp}[g]$ equals $1/4$. This observation coincides with the numerical prediction in \cite{Beisert:2006ez}.

At the strong coupling all the kinematic branch points condense onto $u=\pm 1$. Therefore, if we perform the strong coupling expansion of the generalized scaling function
 \be
   f[g,j]=\frac 1{\e}\(\ f_0+\e f_1+\e^2 f_2+\ldots\ \)
 \ee
 the coefficients $f_n$ should become singular when the end of the root distribution approaches $\pm 1$.

 In the scaling limit with $\ell=\e j$ fixed the position of the branch point at $\e=0$ is given by \cite{Casteill:2007ct,Gromov:2008en}:
 \be
    a[\e=0,\ell]=\frac 12\(b+\frac 1b\),\;\;\;\;b\equiv \sqrt{1+\ell^2}.
 \ee
Therefore, for any positive values of $\ell$ the functions $f_n[\ell]$ are analytic. And also, the large $\ell$ expansion commutes with the strong coupling expansion.  At first few orders the large $\ell$ expansion has BMN-like properties as we mentioned in the introduction.

At $\ell=0$ the functions $f_n[\ell]$ should become singular. To investigate in detail this special case we introduce an additional resolvent $H$ via the relation
\be\label{eq:RH1}
  (D-D^{-1})(R_m-H)=2K_+DR_m
\ee
together with the demand for $R_m$ and $H$ to have the same asymptotics at infinity.

The BES/FRS equation then can be rewritten in the following form\footnote{The equations (\ref{holobes}), (\ref{eq:RH1}) and (\ref{eq:RH2}) were obtained in collaboration with Ivan Kostov and Didina Serban.}
\be\label{eq:RH2}
  (D-D^{-1})(R_m+H)=2K_-DH+2D^{-1}R_h-\frac 2\e\frac{\ell}x\frac{dx}{du}\;.
\ee

Let us consider the region $u^2<1$ and perform an analysis similar to one in \cite{Kostov:2008ax}.
We act by $K_+$ on (\ref{eq:RH1}) and by $K_-$ on (\ref{eq:RH2}) and, since $K_\pm^2=K_\pm$, obtain
 \be
  0&=&K_+\(\;(D+D^{-1})R_m+(D-D^{-1})H\;\),\no\\
  0&=&K_-\(\;(D-D^{-1})R_m-(D+D^{-1})H-2D^{-1}R_h\;\).
 \ee
Treating this expression perturbatively,
we conclude from the definition of $K_\pm$ and after some algebra that for $u^2<1$
\be\label{funnyfunc}
  (D^2+D^{-2})(R_m[u+i0]-R_m[u-i0])=(D-D^{-1})(D^{-1}R_h[u+i0]+DR_h[u-i0]).
\ee
The notion "perturbatively" means that we should perform the strong coupling expansion before evaluating the expression. For instance, the expression $D^{-1}R_h[u+i0]$  should be understood as
\be
  D^{-1}R_h[u+i0]=R_h[u+i0]-i\e\p_u R_h[u+i0]+\ldots
\ee
but not as $D^{-1}R_h[u+i0]=R_h[u-i\e+i0]=R_h[u-i\e]$.

In the region $u^2<a^2$ the discontinuity of $R_m$ is zero by definition, therefore
\be
  D^{-1}R_h[u+i0]+DR_h[u-i0]=0\;.
\ee
At least perturbatively, the zero modes of $D-D^{-1}$ do not contribute. One can show (see, for example \cite{Kostov:2008ax}) that the zero modes of $D-D^{-1}$, even if present, lead to nonperturbative corrections which are not considered in this paper\footnote{However, they are important for the comparison with the O(6) sigma model \cite{Basso:2008tx}.}.

For $u^2>a^2$ the discontinuity of the resolvent $R_h$ is zero.  Therefore, if $a^2<u^2<1$ we conclude from (\ref{funnyfunc}) that
\be
  R_m[u+i0]-R_m[u-i0]=\frac {D^2-D^{-2}}{D^2+D^{-2}}R_h.
\ee
We see that the density of magnons inside the Jukowsky cut $[-1,1]$ has the same magnitude as the density of holes which is of order $j$. For $a<1$ the magnitude of $j$ is of order $g^\a e^{2\pi g(a-1)}$ \cite{Basso:2008tx,Bajnok:2008it} and therefore is exponentially small with respect to the coupling constant. We see that the perturbative expansion of the resolvent in the powers of $\e$ takes place only for $u> 1$.

\section{Perturbative solution at strong coupling}
In the following we will consider the strong coupling limit with $\ell=\e j$ fixed. We assume the following expansion of the resolvents
\be
  R_m&=&\frac 1\e R_{m,0}+R_{m,1}+\e R_{m,2}+\ldots,\no\\
  R_h&=&\frac 1\e R_{h,0}+R_{h,1}+\e R_{h,2}+\ldots.
\ee
We will treat all the equations in this section perturbatively in the sense that first we perform the strong coupling expansion of the resolvents and of the shift operator and then evaluate the expressions.

\subsection{Disappearing of the dressing phase}
As was discussed in the precedent section, in the case when $\ell$ remains finite all the roots are situated outside the Jukowsky cut. We will show that the BES/FRS equation can be considerably simplified in this case.

We start the discussion with the following observation.
Before performing the limit (\ref{introFRS}) the resolvent of the magnons can be represented as
\be\label{mrm}
  \log [M]\,R_m=\suml_k \frac{1}{u-u_k}=\frac 2{1-\frac 1{x^2}}\suml_k \frac 1{x-x_k}+\frac 2{1-x^2}\suml_k \frac 1{\frac 1x-x_k}\;.
\ee
In the continuous limit, the first term on the r.h.s will give a function which has a cut on the interval $(-\infty,-b^*]\cup[b^*,\infty)$ of the Jukowsky plane, where $b^*$ is related to $a$ by
\be
  a=\frac 12\(b^*+\frac 1{b^*}\).
\ee
The second term on the r.h.s of (\ref{mrm}) leads to the function with a cut on $[-1/{b^*},1/{b^*}]$. It is convenient to introduce the resolvent $S[x]$ with the only branch cut $(-\infty,-b^*]\cup[b^*,\infty)$ by the relation
\be\label{rss}
  R_m=S[x]+S[1/x]\,.
\ee

It is important to distinguish the exact position of the branch point $b^*$ with the position of the branch point for $\e=0$ which we denote by $b$.

The equality (\ref{rss}) does not fix S[x] uniquely, but up to an equivalence
\be
  S[x]\simeq S[x]+Q[x]\,,\,Q[x]+Q[1/x]=0\,.
\ee
In particular, $S[x]\simeq \frac 2{1-\frac 1{x^2}}\sum \frac 1{x-x_k}$. The structure of the BES/FRS equation gives us a preferred choice of the representative of \{$S$\}. We demand $(x^2-1)(D-D^{-1})S$ to be analytic everywhere except on the cut $(-\infty,-b]\cup[b,\infty)$. We will refer to this demand as the condition of analyticity\footnote{Before performing the limit (\ref{introFRS}) the resolvent $S[x]$ can be written as $$S[x]=\frac 2{1-\frac 1{x^2}}\frac 1{\log M}\sum\frac 1{x-x_k}+Q[x]$$ with $Q[x]$ being meromorphic odd function with poles at $x=\pm 1$. $Q[x]$ is chosen to satisfy the condition of analyticity.} . Note that the condition of the analyticity should be understood perturbatively in the sense that we explained above. In particular, in the first three orders the condition of analyticity implies that
\be\label{eq:ddm1condition}
  S_0 \;\;\;\;\; \textrm{is analytic outside the cut,} \no\\
  S_1 \;\;\;\;\; \textrm{is analytic outside the cut,} \no\\
  S_2-\frac 16\p_u^2S_0 \;\;\; \textrm{is analytic outside the cut.}
\ee
The preferred choice of the condition of analyticity is explained by the following perturbative equality
\be\label{eq:drm}
  K_+DR_m&=&(D-D^{-1})S[1/x]
\ee
Indeed,
\be
  K_+DR_m&=&\intl_{-1}^1\frac{dv}{2\pi i}\frac{y-\frac 1y}{x-\frac 1x}\frac 1{v-u}(D-D^{-1})(S[y]+S[1/y])=\no\\&=&\oint\frac{dv}{2\pi i}\frac{y-\frac 1y}{x-\frac 1x}\frac 1{v-u}(D-D^{-1})S[1/y]=(D-D^{-1})S[1/x].
\ee
The contour of integration goes clockwise around $[-1,1]$. We used the condition of analyticity to close the contour of integration.

From (\ref{eq:drm}) and (\ref{eq:RH1}) we deduce that
\be\label{hss}
  H&=&S[x]-S[1/x].
\ee
The BES/FRS equation simplifies:
\be\label{eq:besup}
  (1-D^2)S+R_h=-DK_-(D-D^{-1})S+\frac 1\e D\frac {\ell}x\frac {dx}{du}.
\ee
This simplification is quite remarkable. The form of (\ref{eq:besup}) is similar to the BES/FRS equation (\ref{holobes}) with $DR_m$ substituted with $(D-D^{-1})S$ and significant simplification of the kernel which becomes just $K_-$. We see that the convolution of the kernels, which comes from the term with the dressing phase in the Bethe Ansatz, has disappeared at strong coupling. This is an all-loop generalization of the observation, made in \cite{Gromov:2008en}, that the strong coupling expansion of the Bethe equations does not contain $S[1/x]$ terms.

The equation (\ref{eq:besup}) itself cannot be solved since it contains two unknown quantities. Together with it we should use the conjugate equation, which is valid in the lower half plane:
\be\label{eq:besdown}
  (1-D^{-2})S+R_h=-D^{-1}K_-(D-D^{-1})S+\frac 1\e D^{-1}\frac {l}x\frac {dx}{du}.
\ee
If we multiply (\ref{eq:besup}) by $D^{-1}$, (\ref{eq:besdown}) by $D$ and take the difference of the obtained equations on the interval $x^2>b^2$, we will obtain the following result\footnote{Strictly speaking, we obtain $(D-D^{-1})(S[x+i0]+S[x-i0]+R_h)=0$. Asymptotics at infinity assures that it is equivalent to (\ref{eq:SSRh}).}
\be\label{eq:SSRh}
  S[x+i0]+S[x-i0]=-R_h.
\ee
Let us discuss the relation of the equations (\ref{eq:besup}) and (\ref{eq:SSRh}) with the BES equation, which we expect to recover when $R_h=0$ and $\ell=0$. The equation (\ref{eq:besup}) is a variation of the idea to decouple the BES equation into two simpler equations by introducing an additional function \cite{Basso:2007wd,Kostov:2008ax}. From the equalities (\ref{rss}) and (\ref{hss}) we can identify $S$ with the resolvents in \cite{Kostov:2008ax}:
\be
  S[1/x]=R_+,\;\;\;\;S[x]=R_-.
\ee
This identification is possible only perturbatively at strong coupling. We see that at strong coupling the resolvents $R_+$ and $R_-$ are not independent but can be expressed through the one resolvent $S$, the physical meaning of which we discussed above.

 For $R_h=0$ the equation (\ref{eq:SSRh}) simplifies to
\be\label{SSBES}
S[x+i0]+S[x-i0]=0.
\ee
Its solution in the leading order reproduces the AABEK density \cite{Alday:2007qf}. This equation in another parametrization was discussed in \cite{Kostov:2008ax}.

Of course, the equation (\ref{SSBES}) has an infinite number of solutions. The correct solution is chosen from the investigation of the analytical structure of $S$ in the NFS double scaling limit \cite{Basso:2007wd,Kostov:2008ax}.

Note that we obtained the equation $S[x+i0]+S[x-i0]=0$ in a different order of limits than what was used in the strong coupling expansion of the BES equation: we took $\ell\to 0$ after the strong coupling expansion while the BES equation corresponds to the inverse case. Although the order of limits do not affect the equation (\ref{SSBES}) itself, the analysis in the vicinity of the branch point is completely different in two cases. When we apply the NFS double scaling limit we automatically use the order of limits for the strong coupling expansion of the BES equation.

\subsection{The perturbative procedure}
Using the equations (\ref{eq:besup}),(\ref{eq:besdown}), and (\ref{eq:SSRh}), we can perform the strong coupling expansion of the resolvent $S$. The logic of the computations is the following: since $D-D^{-1}=\CO(\e)$, using (\ref{eq:besup}) and (\ref{eq:besdown}), we can express $R_{h,n}$ in terms of $S_{m}$ with $m<n$. Then, by solving the Riemann-Hilbert problem (\ref{eq:SSRh}) for $S_n$, we can express $S_n$ in terms of $R_{h,n}$. The final result is given by the equations (\ref{eq:finalrecursion}) which allows to express $S_n$ in terms of $S_{m}$ with $m<n$.

In the following we discuss the details of the calculations which lead to (\ref{eq:finalrecursion}).

The equation (\ref{eq:besup}) allows us to find $R_h[u]$ in the upper half plane. Let us introduce the function $\Phi$ by the following relation
\be\label{Phiup}
  \Phi=D^{-1}R_h-\frac 1\e \frac {\ell}{x}\frac {dx}{du},\;\;\Im[u]>0.
\ee
From (\ref{eq:besup}) we conclude that we can rewrite the function $\Phi$ in the following way
\be\label{Phi}
\Phi[x]=(D-D^{-1})S[x]-\intl_{-1+i0}^{1+i0}\frac{dv}{2\pi i}\frac{y-\frac 1y}{x-\frac 1x}\frac 1{v-u}(D-D^{-1})(S[y]-S[1/y])=\no\\=
(D-D^{-1})S[x]+\intl_\G\frac {dy}{2\pi i}\(\frac{y-\frac 1y}{x-\frac 1x}\)^2\(\frac 1{y-x}-\frac 1{y-\frac 1x}\)(D-D^{-1})S[1/y]\;,
\ee
where $\G=\G_++\G_-$, $\G_\pm$ - the upper/lower unit semicircle from $-1$ to $1$.

We perform the deformation of the contour of integration, shrinking it towards the interval $[-1,1]$. Since $\Im[u]>0$, the contour $\G_-$ passes through the point $1/{x[u]}$. The residue at this point cancels the first term in the r.h.s of (\ref{Phi}) and we obtain
\be\label{eq:phi}
  \Phi[x]&=&\intl_\gamma\frac{dy}{2\pi i}\(\frac{y-\frac 1y}{x-\frac 1x}\)^2\(\frac 1{y-x}-\frac 1{y-\frac 1x}\)(D-D^{-1})S[1/y],\no\\
  \g&=&\g_++\g_-,\;\;\g_\pm=[-1\pm i0,1\pm i0]\;.
\ee
We take (\ref{eq:phi}) as the definition of $\Phi$ for any values of $x$. From the equation (\ref{eq:besdown}) we get that
\be\label{Phidown}
  \Phi=D\,R_h-\frac 1\e \frac {\ell}{x}\frac {dx}{du},\;\;\Im[u]<0.
\ee
As it follows from (\ref{eq:phi}), the function $\Phi[x]$ is discontinuous everywhere on the real axis of the Jukowsky plane. However, from (\ref{Phidown}) and (\ref{Phiup}) we can expect that
\be\label{contphi}
  D^{-1}\Phi[x-i0]=D\Phi[x+i0],\;\;\; x^2>b^2,
\ee
since $R_h$ should be continuous on the interval $x^2>b^2$. Indeed, we can prove the equality (\ref{contphi}) by induction using the definition (\ref{eq:phi}) of $\Phi$.

Using (\ref{Phiup}),(\ref{Phidown}), and (\ref{contphi}) we can reformulate (\ref{eq:SSRh}) as an equation for the function $S$:
\be\label{Sphi}
  S[x+i0]+S[x-i0]=-\frac 1{D+D^{-1}}\(\Phi[x+i0]+\Phi[x-i0]+\frac 1\e\frac {4\ell}{x-\frac 1x}\),\;\;\; x^2>b^2.
\ee
The general solution of (\ref{Sphi}) which respects the cut structure of $S$ is the following:
\begin{small}
\be\label{eq:finalrecursion}
S_n&=&S_{nh,n}+S_{hom,n}-\[\frac 1\e\frac 2{D+D^{-1}}\]_n\frac{\ell}{x-\frac 1x},\no\\
 S_{hom,n}&=&x\sqrt{b^2-x^2}\(\suml_{k=1}^{2n+1}\frac{a_{n,k}}{(1-x^2)^k}+\suml_{k=1}^{n}\frac{c_{n,k}}{(b^2-x^2)^k}\)\;,\no\\
S_{nh,n}&=&I_{1,n}+I_{2,n},\no\\
  I_{1,n}&=&\[-\frac 1{D+D^{-1}}\intl_\g \frac {dy}{2\pi i}\(\frac{y-\frac 1y}{x-\frac 1x}\)^2(D-D^{-1})S[1/y]\times\right.\no\\&\times&\left.\(\frac 1{y-x}\(1-\frac{\sqrt{b^2-y^2}}{\sqrt{b^2-x^2}}\)-\frac {1}{y-\frac 1x}\(1-\frac{\sqrt{b^2-\frac 1{y^2}}}{\sqrt{b^2-x^2}}\)\)\]_n,\no\\
  I_{2,n}&=&\[\frac 1{D+D^{-1}}\intl_\b \frac {dy}{2\pi i}\(\frac{y-\frac 1y}{x-\frac 1x}\)^2\frac {1}{y-\frac 1x}\frac{\(b^2-\frac{1}{y^2}\)^{n-1/2}}{\(b^2-x^2\)^{n-1/2}}(D-D^{-1})S[1/y]\]_n.
\ee
\end{small}

Here $\b=\b_++\b_-$ with $\b_\pm=[-1/b\pm i0,1/b\pm i0]$.

We remind that we assume the perturbative expansion of $S$
\be
  S=\frac 1\e\(\ S_0+\e S_1+\e^2 S_2+\ldots\ \).
\ee
By the notation $[F]_n$ we understand the coefficient in front of $\e^{n-1}$ of the perturbative expansion of $F$ in powers of $\e$. The perturbative expansion of the integrals $I_{1}$ and $I_{2}$ is understood in the following way: first we expand the resolvent and the expressions containing the shift operator $D$ and then perform the integration. Since $D-D^{-1}=\CO(\e)$, the r.h.s of (\ref{eq:finalrecursion}) contains only $S_m$ with $n<m$. Therefore, the solution (\ref{eq:finalrecursion}) defines the recursive procedure which allows to express $S_n$ in terms of $S_m$ with $n<m$.

The constants $a_{n,k}$ are fixed by the condition of analyticity, which in the first three orders is given by (\ref{eq:ddm1condition}).

To fix the coefficients $c_{n,k}$ we need an additional information.
At the first two orders these coefficients are fixed by the following properties:
\bi
  \item $S_n$ has at most a $\(b^2-x^2\)^{1/2-n}$ singularity at $x=\pm b$ (in particular, $S_1[\pm b]=0$),
  \item $S_1$ behaves as $-\frac i\e$ for $x\rightarrow \infty+i0$,
  \item $S_n$ decreases at infinity for $n>1$.
\ei
The asymptotics at infinity fixes $c_{n,1}=-a_{n,1}$ for any $n$.

These conditions are not sufficient to fix the coefficients $c_{n,k}$ with $n>2$ and $k>1$ which appear starting with two loops. The analysis which allows to fix these coefficients is given in Sec.~\ref{nfsforfrs}.

 The generalized scaling function, determined by the formula (\ref{f2cj}), can be expressed in terms of the coefficients $a_{n,k}$ and $c_{n,k}$ by
\be
 f_{n}=-b\(\sum_{k=1}^{2n+1}a_{n,k}+\sum_{k=1}^{n}\frac{c_{n,k}}{b^{2k}}\)-\ell\,\delta_{n,0}.
\ee
In the following we will focus on the first three orders.

\subsection{Tree and one loop level}
The leading order solution follows easily from (\ref{eq:finalrecursion}):
\be\label{treelevelsolution}
  S_0=-\frac 1{x-\frac 1x}\({a_{0,1}\sqrt{b^2-x^2}}+\ell\).
\ee
The asymptotics at infinity fixes $a_{0,1}=-1$ and the condition of analyticity is satisfied only if
\be
  b=\sqrt{\ell^2+1}\,.
\ee
We obtain
\be
  f_0=\sqrt{\ell^2+1}-\ell.
\ee
At the one loop level the solution (\ref{eq:finalrecursion}) reads
\be
  S_1&=&S_{nh,1}+S_{hom,1},\no\\
  S_{hom,1}&=&x\sqrt{b^2-x^2}\(\frac {a_{1,1}}{1-x^2}+\frac {a_{1,2}}{(1-x^2)^2}-\frac {a_{1,1}}{b^2-x^2}\),\no\\
  S_{nh,1}&=&I_{1,1}+I_{2,1},\no\\
  I_{1,1}&=&\(\frac{2}{x-\frac 1x}\)^2\intl_\g\frac {dy}{2\pi}\;\frac{y^2-1}2\(-\frac {d}{dy}S_0[1/y]\)\times\no\\&&\times\(\frac 1{y-x}\(1-\frac{\sqrt{b^2-y^2}}{\sqrt{b^2-x^2}}\)-\frac {1}{y-\frac 1x}\(1-\frac{\sqrt{b^2-\frac 1{y^2}}}{\sqrt{b^2-x^2}}\)\),\no\\
   I_{2,1}&=&-\(\frac{2}{x-\frac 1x}\)^2\intl_\b\frac {dy}{2\pi}\;\frac{y^2-1}2\(-\frac {d}{dy}S_0[1/y]\)\frac {1}{y-\frac 1x}\frac{\sqrt{b^2-\frac 1{y^2}}}{\sqrt{b^2-x^2}}.
\ee
The integrals $I_{1,1}$ and $I_{2,1}$ can be evaluated explicitly (see appendix \ref{sec:dispersion} for the details) and the coefficients $a_{1,1}$, $a_{1,2}$ are fixed by the condition of analyticity of $S_1$ at $x=\pm 1$. The resolvent $S_1$ is given by the following expression:
\begin{small}
\be\label{eq:s1loop}
&S_{nh,1}&=-2\ell(\log[b-1]-2\log[b]+\log[b+1])+\no\\&+&\!\!\!\!\!\!\!\!\frac {\ell^2}{\sqrt{b^2-x^2}}\(4\log[b]-2\log[b+1]-\log[b^2+1]\)+\no\\&+&\!\!\!\!\!\!\!\!\ell\(x+\frac 1x\)\log\[\frac{(1-x)(\sqrt{b^2-x^2}+\ell\ x)}{(1+x)(\sqrt{b^2-x^2}-\ell\ x)}\]+\no\\&+&\!\!\!\!\!\!\!\!\frac{\(\frac{b^2}{2}(x+\frac 1x)-x\)}{\sqrt{b^2-x^2}}\log\[\frac{(1-x)(b+x)(\sqrt{b^2-x^2}+\ell\ x)}{(1+x)(b-x)(\sqrt{b^2-x^2}-\ell\ x)}\]+\no\\&+&\!\!\!\!\!\!\!\!\!\!\frac{\(\frac {b^2}{2}(x+\frac 1x)-\frac 1x\)}{\sqrt{b^2-\frac 1{x^2}}}\log\[\frac{\left(x\sqrt{b^2-\frac 1{x^2}}-\sqrt{b^2-x^2}\right)\!\left(x\sqrt{b^2-\frac 1{x^2}}+\ell\right)}{\left(x\sqrt{b^2-\frac 1{x^2}}+\sqrt{b^2-x^2}\right)\!\left(x\sqrt{b^2-\frac 1{x^2}}-\ell\right)}\]\!\!,
\ee
\be
a_{1,1}&=&\frac 1{\pi \ell^4}(3 b^3-3 b^2+b-1-4b^2 \left(b^2-3\right) \log[b] +2b^2 \left(b^2-3\right) \log[b+1] +\no\\&&+\left(b^2-1\right)^2 \log[b-1]+\frac{1}{2} \left(b^4-4 b^2-1\right) \log [b^2+1])\;,\no\\
a_{1,2}&=&\frac 1{\pi\ell^2}\(\,4-4 b-8 \log[b]+4 \log[b+1]+2 \log[b^2+1]\,\)\;.
\ee
\end{small}
The one-loop generalized scaling function is given by
\begin{footnotesize}
\be
f_1=\frac{b-1+8 b^2 \log[b]-(b^2+1) \log[b^2+1]-2(b^2-1) \log[b-1]-2 b^2 \log[b+1]}{2\pi\;b}
\ee
\end{footnotesize}
The results for $f_0[\ell]$ and $f_1[\ell]$ coincides with the results in \cite{Casteill:2007ct,Belitsky:2007kf} and therefore with the perturbative calculations in the string theory. The difference between $S_1$ and the resolvent found by Belitsky \cite{Belitsky:2007kf} is an odd meromorphic function of $x$ with poles at $x=\pm 1$. This difference does not contribute to $R_m$.

\subsection{Two loops}
The two loop resolvent is given by the solution (\ref{eq:finalrecursion}) with $n=2$.
The computation at this order is more difficult since we have to fix the coefficient $c_{2,2}$ in $S_{hom,2}$ which requires an additional analysis of the initial equation (\ref{holobes}). We postpone this analysis to the next section. The result is the following: the resolvent $S_2$ is singular at the point $x=b$ with the leading square root singularity given by
\be\label{s2sing1}
  S_2&=&\frac{b^3}{(2b)^{3/2}}\frac {\tilde \CQ}{(b-x)^{3/2}}+\ldots
\ee
with
\be\label{tildeCQ}
   \tilde \CQ&=&-\frac{2b^4}{\pi^2\ell^6}\(\Theta(\Theta-4)+\frac 23\pi^2\),\no\\
   \Theta&=&\frac 1{4 b^3}(-2+2 b-6 b^2+6 b^3-4b^3\log[2]-4b^2\left(2 b^2+b+2\right)\log[b]+\no\\&&+2\left(b^4-2 b^3-2 b^2+1\right) \log[b-1]+\left(4 b^4+4 b^3+4 b^2\right) \log[b+1]+\no\\&&+\left(b^4+4 b^2-1\right) \log[b^2+1]).
 \ee
 This singularity comes from $S_{hom,2}$ and $S_{nh,2}$. The contribution from these two terms is the following
 \be\label{s2sing2}
  S_{hom,2}&=&\frac{b}{(2b)^{3/2}}\frac{c_{2,2}}{(b-x)^{3/2}}+\ldots,\no\\
  S_{nh,2}&=&\frac{b^3}{(2b)^{3/2}}\frac{P}{(b-x)^{3/2}}+\ldots,
 \ee
where $P$ is a number given by a complicated integral. We give the explicit form of $P$ in the appendix \ref{sec:2loop}.

Comparing (\ref{s2sing1}) with (\ref{s2sing2}) we get the following expression for the constant $c_{2,2}$:
\be
  c_{2,2}=b^2(\tilde \CQ-P).
\ee
The form of the solution suggests that in the vicinity of $x=1$ the function $S_{nh,2}$ can be expanded as
\be
  S_{nh,2}=\frac {I_{s2}}{(x-1)^2}+\frac{I_{s1}}{(x-1)}+\CO(1).
\ee
The coefficients $a_{2,k}$ are fixed from the condition of analyticity which  implies that
\be
  S_{hom,2}+\frac {I_{s2}}{(x-1)^2}+\frac{I_{s1}}{(x-1)}-\frac 12\frac {d^2}{du^2}\(\frac \ell{x-\frac 1x}+\frac 13 S_0\)
\ee
should be analytic at $x=1$.

Collecting all the coefficients together, we get the following expression for the 2-loop generalized scaling function
\be\label{2loopexp}
  f_2^{FRS}[\ell]=\frac 1{\sqrt{1+\ell^2}}\(P-\tilde \CQ-2\ell I_{s1}+4\ell I_{s2}\(1+\frac 1{2\ell^2}\)-\frac {31}{24}\frac 1{\ell^6}-\frac 73\frac 1{\ell^4}-\frac 1{\ell^2}\).
\ee
Explicit expression for $-2\ell I_{s1}+4\ell I_{s2}\(1+\frac 1{2\ell^2}\)$ is given in the appendix \ref{sec:2loop}.

The numerical comparison with the result of \cite{Gromov:2008en} shows that the scaling function (\ref{2loopexp}) and the scaling function obtained from the Bethe Ansatz in the order of limits (\ref{stringlimit}) are related as
\be\label{f2loop}
  f_2^{FRS}[\ell]&=&f_2^{BA}[\ell]+\d[\ell],\\\d[\ell]&=&\frac 1{\sqrt{1+\ell^2}}\(\frac 1{24}\frac 1{\ell^6}+\frac 1{12}\frac 1{\ell^4}\).\no
\ee

The expression (\ref{2loopexp}) can be expanded for the large values of $\ell$. We are interested in the leading term from which we can find the coefficient $c_{12}$ defined in (\ref{BMNlike}). Evaluating first
\be
  \tilde \CQ&=&\(-\frac 43+\frac{8}{\pi^2}-2\frac{(\log[2]+\log[\ell])\log[2\ell]}{\pi^2}\) \frac 1{\ell^{2}}+\CO(\ell^{-4}),\no\\
  P&=&\(6-\frac {32}{\pi^2}-2\frac{(\log[2]+\log[\ell])\log[2\ell]}{\pi^2}\) \frac 1{\ell^{2}}+\CO(\ell^{-4}),\no\\
  -2\ell I_{s1}+4\ell I_{s2}\(1+\frac 1{2\ell^2}\)&=&\(-6+\frac {40}{\pi^2}\)\frac 1{\ell^{2}}+\CO(\ell^{-4}),
\ee
we obtain
\be\label{largel}
  f_2[\ell]=\frac 13\frac 1{\ell^3}+\CO(\ell^{-5})\;\;\;\textrm{and}\;\;\;c_{12}=\frac {16}3.
\ee
The term $\d[{\ell}]$ does not contribute to the coefficient $c_{12}$. To verify (\ref{largel}), we performed numerically large $j$ expansion of the generalized scaling function at $g=0$. The details of the computation can be found in the appendix~\ref{sec:largej}.

\section{Behavior of the solution near the branch point.}\label{nfsforfrs}
In this section we explain how to obtain the coefficient $\tilde \CQ$ in (\ref{s2sing1}).

The perturbative solution (\ref{eq:finalrecursion}) is defined for the values of $x$ such that $|x-b|$ is much larger than $\e$. At the point $b$ the perturbative expansion is not valid and the solution (\ref{eq:finalrecursion}) develops a singularity.
Let us understand what type of the singularity is expected at this point. For this we take the difference of (\ref{eq:besup}) and (\ref{eq:besdown}) for $x^2>b^2$, getting\footnote{Strictly speaking, we get $(D-D^{-1})(\ref{eq:cont})=0$. The equation (\ref{eq:cont}) is deduced from the conditions of decreasing at infinity}
\be\label{eq:cont}
  D\,S[x+i0]+D^{-1}S[x-i0]=K_-(D-D^{-1})S-\frac 1{\e}\frac {2\ell}{x-\frac 1x}.
\ee
The r.h.s of this equation is analytic at $x=b$ and therefore we can write the following equality:
\be\label{eq:ssing}
  D\,S_{sing}[x+i0]+D^{-1}S_{sing}[x-i0]=0,
\ee
where by $S_{sing}$ we denoted a part of the resolvent $S$ which is singular at $x=b$. We will be only interested in the leading singularities of $S$ at each order of the perturbation theory. At tree and one loop level they can be found directly from the corresponding solutions (\ref{treelevelsolution}) and (\ref{eq:s1loop}). We find the most general form of the two loop singularity by solving the equation (\ref{eq:ssing}). Combining the leading tree, one and two loop singularities together, we get the following expression:
\be\label{eq:qprediction}
 S_{sing}&=&\frac{\sqrt{b-x}}{\e}\frac{\sqrt{2b^3}}{\ell^2}\(1-\frac{\e b^2}{\pi\ell^2}\frac{\log[b-x]+\Theta}{b-x}-\right.\no\\&&\left.-\frac 12\(\frac {\e b^2}{\pi\ell^2}\)^2\frac {\CQ-4(\log[b-x]+\Theta)+(\log[b-x]+\Theta)^2}{(b-x)^2}+\ldots\),\no\\
 \Theta&=&\frac 1{4 b^3}(-2+2 b-6 b^2+6 b^3-4b^3\log[2]-4b^2\left(2 b^2+b+2\right)\log[b]+\no\\&&+2\left(b^4-2 b^3-2 b^2+1\right) \log[b-1]+\left(4 b^4+4 b^3+4 b^2\right) \log[b+1]+\no\\&&+\left(b^4+4 b^2-1\right) \log[b^2+1]).
\ee
The coefficient $\CQ$ is arbitrary. To fix $\CQ$ we need to consider the BES/FRS equation in the double scaling
regime which is defined as follows. We introduce the variable
\be\label{zdoublescaling}
  z=\frac{u-a}{\e}\;
\ee
and perform the double scaling limit $\e\to 0$ with $z$ fixed. Note that $a$ is the exact position of the branch point.

Since $\e\gg d$, the treatment of the resolvent as the analytic function of $z$ with cuts is valid in the double scaling limit. To compare, if we apply the double scaling limit for the order of limits (\ref{stringlimit}), used for the perturbative calculations in the string theory, we will see the separate poles of the resolvent since $\e\ll d$. We illustrate the discussion in this paragraph in Fig.~2.
\begin{figure}\label{fig:sccales}\centering
\includegraphics[width=4cm]{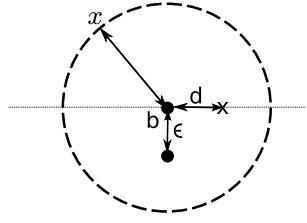}\qquad
\caption{The scales that appear in the Bethe equations. $b$ is the position of the smallest positive Bethe root, $d$ is the distance between two neighboring roots, $\e$ appears as the shift in (\ref{baxtercomplete}).}
\end{figure}

For our purposes it is sufficient to keep only the leading order of the double scaling limit. From the solution (\ref{eq:finalrecursion}) we see that the leading order scales as $\e^{-1/2}$. The functions analytic at the point $u=a$ scale with integer powers of $\e$ and therefore will not appear in the double scaling limit.

We see that the equations (\ref{eq:besup}) and (\ref{eq:besdown}) reduce to the following ones
\be\label{doublescalingecuation}
  S[z]-S[z+2i]+R_{h}[z]&=&0\;\;\;\textrm{for u.h.p},\no\\
  S[z]-S[z-2i]+R_{h}[z]&=&0\;\;\;\textrm{for l.h.p}.
\ee
 To solve these equations we perform the inverse Laplace transform. We define the inverse Laplace transform for $S$ and $R_h$ in the following way:
\be\label{ilt}
  \tilde S[s]=\intl_{i\infty-0}^{-i\infty-0}\frac {dz}{2\pi i}S[z]e^{z s}, s<0,\no\\
  \tilde R_h[s]=\intl_{-i\infty+0}^{-i\infty+0}\frac {dz}{2\pi i}R_h[z]e^{z s}, s>0.
\ee
We chose different contours of integration for $S$ and $R_h$ in order to avoid the cuts of the resolvents. The functions $\tilde S[s]$ and $\tilde R_h[s]$ are defined for any complex value of $s$ by analytical continuation of (\ref{ilt}). The cut of $S[z]$ implies the existence of the cut for $\tilde S[s]$ which we define to be on the ray $s>0$. Equivalently, there is a cut for $\tilde R_h[s]$ which we define to be on the ray $s<0$.

The inverse Laplace transform of the equations (\ref{doublescalingecuation}) gives the following equations valid for $s>0$:
\be\label{laplacebes}
  (1-e^{-2is})\tilde S[s-i0]+\tilde R_h[s]&=&0,\no\\
  (1-e^{+2is})\tilde S[s+i0]+\tilde R_h[s]&=&0.
\ee
They imply the equation on $\tilde S[s]$
\be
  \tilde S[s+i0]=-e^{-2is}\tilde S[s-i0],\;\;\;s>0\;,
\ee
which is solved by
\be\label{Q}
  \tilde S[s]=\frac {\G[1-\frac s\pi]}{(-s)^{3/2}}e^{-\frac s\pi+\frac s\pi\log[-\frac s\pi]}A[s].
\ee
Here $A[s]$ is a function with trivial monodromy. The factor $\Gamma[1-\frac s\pi]e^{-\frac s\pi}$ can be absorbed in $A[s]$. The reason to keep it explicitly will be clear below.

At large $z$ the function $S[z,\e]$ should be glued with the solution (\ref{eq:finalrecursion}). In particular, the leading $\e\to 0$ order of $S[z,\e]$ in the double scaling limit should reproduce (\ref{eq:qprediction}). This demand is satisfied if $A[s]$ is expanded in positive integer powers of $s$ around $s=0$:
\be\label{qsmallsexpansion}
  A[s]=\a(1+\a_1 s+\a_2 s^2+\ldots).
\ee
Indeed, assuming (\ref{qsmallsexpansion}) we get from the Laplace transform of $\tilde S[s]$ the following large $z$ expansion for $S[z]$:
\be\label{eq:expansionz}
S[z]=\a\frac{\sqrt{z}}{\e}\(\ \suml_{n=0}^\infty \frac {d_{0,n}}{z^n}+\!\suml_{n=1}^\infty \frac {d_{1,n}\log[z]}{z^n}+\!\suml_{n=2}^\infty \frac {d_{2,n}(\log[z])^2}{z^n}+\ldots\).
\ee
The coefficients $d_{0,n}$ are expressed in terms of  $\a_m$ with $m\leq n$ and $d_{k,n}$ are expressed in terms of $d_{0,m}$ with $m\leq n-k$. Upon identification\footnote{We remind that $b^*$ is the exact position of the branch point of the resolvent $S[x]$ and $b\equiv\sqrt{1+\ell^2}$.}
\be
  z=\frac 12\frac {\ell^2}{b^2}\frac{x-b^*}{\e},\;\;\a=\frac i{\sqrt{\pi b \e}}\frac {b^3}{\ell^3},\;\;b^*=b+\frac{\e}{\pi}(\Theta-2\log 2-\log\pi)+\CO(\e^2)
\ee
the expression (\ref{eq:expansionz}) coincides with (\ref{eq:qprediction}) with
\be\label{CQ}
  \CQ=\pi^2\(\frac 23-\a_1^2+2\a_2\).
\ee
We used the substitution $z=\frac {b-\frac 1b}{2b}\frac{x-b^*}{\e}$ instead of (\ref{zdoublescaling}) which is possible for the consideration of only the leading singularities. The overall normalization $\a$ is chosen to fit (\ref{eq:expansionz}) with the singularity of $S_0$. The exact position of the branch point $b^*$ is chosen to fit (\ref{eq:expansionz}) with the singularity of $S_1$. The coefficients $\a_1$ and $\a_2$ are still arbitrary. They define the unknown coefficient $\CQ$ through (\ref{CQ}).

To fix $\a_1$ and $\a_2$ we use the fact that $S[z]$ has a logarithmic singularity at the origin. Therefore, $S[z]$ is expanded\footnote{the expansion of the type $\tilde r_0+\tilde r_1 z+\tilde r_2 z^2$ may contribute as well, but only in subleading orders.} at the origin as
\be\label{logexpansionRh}
  S[z]=(r_0+r_1 z+r_2 z^2+\ldots)\log[-z]\;.
\ee
This type of the singularity comes from the logarithmic singularity of the resolvent at finite coupling. Another way to see this singularity is to take the difference of the first and the second equations in (\ref{doublescalingecuation}) for $z>0$. The resulting equation is
\be
  S[z+i0]-S[z-i0]=S[z+2i]-S[z-2i].
\ee
We see that the discontinuity of $S[z]$ is given by the function $S[z+2i]-S[z-2i]$ which is analytic at $z=0$. This implies the expansion (\ref{logexpansionRh}) for the function $S[z]$.

The series (\ref{logexpansionRh}) has finite radius of convergence due to the branch points $z=2i\MZ$ of $S[z]$ in the nonphysical plane. This means that $\tilde S[s]$ admits only an asymptotic expansion at infinity
\be\label{eq:logrz}
  \tilde S[s]=-\frac{r_0}s+\frac {r_1}{s^2}-2\frac {r_2}{s^3}+\ldots
\ee
which is valid in the cone $\vartheta\leq\arg s\leq 2\pi-\vartheta$ with arbitrarily small $\vartheta$. In the same cone the asymptotic expansion of $A[s]/\tilde S[s]$ can be performed
\be
  \frac{(-s)^{3/2}}{\Gamma[1-\frac s\pi]}e^{\frac s\pi-\frac s\pi\log[-\frac s\pi]}=\frac s{\sqrt{2}}\(1+\frac 1{12}\frac \pi s+\frac 1{288}\frac{\pi^2}{s^2}+\ldots\).
\ee
Therefore, $A[s]$ in this cone also has an asymptotic expansion
\be\label{eq:asexp}
  A[s]=q_0+\frac{q_1}s+\frac {q_2}{s^2}+\ldots.
\ee
On the other hand, using (\ref{laplacebes}), we can express $A[s]$ in terms of $R_h$:
\be\label{definitionofQ}
  A[s]=-\frac 1{2 \pi}\G\[\frac s\pi\]\,s^{3/2}\,e^{\frac s\pi-\frac s\pi\log[\frac s\pi]}\, \tilde R_h[s].
\ee
Repeating the same arguments we conclude that $A[s]$ admits asymptotic expansion (\ref{eq:asexp}) in the cone $-(\pi-\vartheta)\leq\arg s\leq\pi-\vartheta$ with arbitrarily small $\vartheta$. Therefore, the series (\ref{eq:asexp}) is valid everywhere providing with this to be convergent.

Since $\tilde R[s]$ is analytic outside $s<0$ and $\tilde S[s]$ is analytic outside $s>0$, from the comparison of (\ref{definitionofQ}) and (\ref{Q}) we deduce that $A[s]$ is analytic in $\MC^*$.

This allows us to identify\footnote{we assume the convergence of the series (\ref{qsmallsexpansion}).} (\ref{eq:asexp}) with (\ref{qsmallsexpansion}). We conclude that $A[s]$ is a constant and therefore
\be\label{CQfinal}
  \CQ=\frac 23\pi^2.
\ee
From (\ref{eq:qprediction}) and (\ref{CQfinal}) we get the expression (\ref{tildeCQ}) for $\tilde\CQ$.

\section{Conclusions}
In this paper we computed the strong coupling expansion of the generalized scaling function form the BES/FRS equation up to two loops. At the two-loop level the result is different from what was obtained in \cite{Gromov:2008en}. A possible reason for the discrepancy is in the different order of limits that was used in two approaches. The BES/FRS equation is derived in the limit (\ref{introFRS}) in which we can neglect one of the terms in the l.h.s of the equation (\ref{baxtercomplete}). On the other hand, in the order of limits (\ref{stringlimit}) used in \cite{Gromov:2008en}, this is not justified near the branch point of the resolvent. The order of limits is certainly not important for the calculations at the tree and the one-loop level since the analysis in the vicinity of the branch point is needed starting from the two loops.

We also gave a prediction for the leading term of the large $\ell$ expansion of $f_2[\ell]$ (\ref{largel})  and checked it numerically using the BMN-like properties of the expansion (\ref{BMNlike}).  In view of the discrepancy between $f_2^{BA}[\ell]$ and $f_2^{string}[\ell]$ it would be interesting if one can reproduce the result (\ref{largel}) from string theory calculations.

\bigskip
\leftline{\bf Acknowledgments}

\noindent The author thanks to B. Basso, M.Beccaria, A.Belitsky, N.Gromov,  G. Korchemsky, A.Tseytlin and especially to I.Kostov and D.Serban for many useful discussions. The author thanks to I.Kostov and D.Serban for collaboration in the initial stages of the project. This work has been supported by the  European Union
through ENRAGE network (contract MRTN-CT-2004-005616).

\appendix

\section{The structure of the resolvent in the vicinity of the branch point}
\label{appendixa}
Our aim is to show that the resolvent in the vicinity of the branch point has the logarithmic cut. For the sake of simplicity we discuss the Bethe eqautions for $g=0$. We choose the normalization of $u$ such that the Bethe equations at zero coupling constant have the form
 \be
  \(\frac{u_k+\frac i2}{u_k-\frac i2}\)^{L}=\prodl_{\substack{j=1 \\j\neq k}}^{M}\frac{u_k-u_j-i}{u_k-u_j+i}\;.
 \ee
 If we take the logarithm of the Bethe equations, we obtain
 \be\label{eq:force}
  L\;F[2u_k]&+&\suml_{\substack{j=1 \\j\neq k}}^{M} F[u_k-u_j]=\sign[u_k]\;,\\
  F[u]&=&\frac 1{2\pi i}\log \frac{u+i}{u-i}\;.\no
 \ee
 The equation ($\ref{eq:force}$) can be interpreted as the force equilibrium equation in classical mechanics. For the distances between the particles much larger than one, the interaction between the particles can be approximated with the Coulomb force $F[u_k-u_j]\simeq \frac 1{\pi}\frac 1{u_k-u_j}$. In this case the density of the particles is approximated by the square root cut in the leading order of the large spin limit.

 However, for the scaling considered in the current paper we have the opposite situation: the distances between the particles with the smallest absolute values of rapidities are much smaller than one. Therefore, the Coulomb approximation is not applicable. To describe the distribution of roots in the considered limit it is better to represent $F[u]$ as
 \be
  F[u]=-\frac 1{\pi}\arctan[u]+\frac 12\sign[u]\;.
 \ee
 If we introduce the effective force $F_{\textrm{eff}}[u]=L\, F[2u]-\suml_{\substack{j=1 \\j\neq k}}^{M}\frac 1{\pi}\arctan[u-u_j]$, then for the positive roots equation of the equilibrium will be written as
 \be
    F_{\textrm{eff}}[u_{M/2+k}]=\frac 12+k
 \ee
 Since $F_{\textrm{eff}}$ is a smooth function, we immediately get that in the vicinity of the branch point
 \be
  u_k-u_1\simeq \frac {k}{F^\prime[u_{M/2}]}\simeq \frac {k-1}L
 \ee
 The last estimation comes from the dominant $L\,F[2u]$ term of the $F_{\textrm{eff}}$ and is consistent with assumption of the small distance between the roots. It is valid for $u_{M/2}\ll L$ which is the case for any finite $j$. Equidistant distribution between the roots corresponds to the logarithmic type of the branch point in the continuous limit.

 The situation does not change for any finite value of the coupling constant. The additional terms in the Bethe Ansatz do not change the arguments used in derivation because the interaction introduced by them is nonlocal for any finite $g$.

 At the infinite value of the coupling constant two effects appear. First, the roots scale generically as $g$ and therefore the distance between them becomes large. In fact, in the order of limits (\ref{stringlimit}) used for the perturbative calculations in the string theory, the branch point changes into a square root type for any nonzero $\ell$.

 Second, at strong coupling and for $\ell\to 0$ the distribution of roots approaches the Jukowsky branch point, where the additional\footnote{to the $g=0$ Bethe equation} terms of the Bethe equation become local. They result in the change of square root cut into the $u^{-1/4}$ behavior of the resolvent.

\section{Large $j$ expansion}\label{sec:largej}
Large $j$ expansion for the Bethe equations (\ref{sl2bethe}) at $g=0$ was done numerically in the regime (\ref{introFRS}) \cite{Beccaria:2008nf}. However, only the first two terms of this expansion were given. We need the third term in order to verify the prediction for the coefficient $c_{12}=\frac {16}3$.

The generalized scaling function can be found if we know the density of holes (\cite{Freyhult:2007pz,Beccaria:2008nf}):
\be\label{energy}
  f(g,j)=8g^2+2g^2\int_{-a}^{a} du\(\psi\(\frac 12+iu\)+\psi\(\frac 12-iu\)-2\psi(1)\)\rho_h(u)+\CO(g^4).\no\\
\ee
The density of holes satisfies the integral equation\footnote{it is derived under the same assumptions as the BES/FRS equation.}
\be\label{eq:1rootrho}
  \rho_h\!=\frac{2}{\pi}-\frac j{2\pi}\(\psi[\frac 12+i u]+\psi[\frac 12-iu]\)\!\!+\!\!\intl_{-a}^a\!\frac {dv}{2\pi}\!(\psi[i(u-v)]\!+\!\psi[-i(u-v)])\rho_h[v]
\ee
and is normalized by
\be\label{normalization}
  j=\int_{-a}^a \rho_h(v) dv.
\ee
Substituting the normalization condition into (\ref{eq:1rootrho}) we get an integral equation which depends only on the parameter $a$. Solving numerically this equation and fitting the results for $j$ in the range from 30 to 150, we obtain the following large $j$ expansion:
\be
  f(g,j)=g^2(\frac{8.0000}{j}-\frac {6.79}{j^2}+\frac{5.33}{j^3}-...)+\CO(g^4),
\ee
which is consistent with (\ref{largel}).

\section{Evaluation of the Cauchy type integrals}\label{sec:dispersion}
The solution (\ref{eq:finalrecursion}) suggests evaluation of the Cauchy type integrals. This evaluation can be simplified if we will use the observation explained below.

Suppose that we know the value of the following Cauchy type integral
\be
  I[x]=\int_\g dz \frac {f[z]}{z-x}
\ee
with $\g$ - any contour, closed or not. Then for any rational function $Q[x]$, regular on the contour
\be\label{f}
  F[x]\equiv\int_\g dz \frac{Q[z]f[z]}{z-x}=Q[x]I[x]+R[x]
\ee
with $R[x]$ being the rational function.

Indeed, discontinuous part of F is given by.
\be
  F[x+i0]-F[x-i0]=Q[x]f[x]=Q[x](I[x+i0]-I[x-i0]).
\ee
The most general solution of the last equation is
\be
  F[x]=Q[x]I[x]+R[x]
\ee
with $R[x]$ being a rational function. Since F[x] should be the value of the integral (\ref{f}), it should be analytic outside the contour $\g$ and decrease at infinity. $R[x]$ is chosen to fulfill this properties.

 We use it to compute integrals in (\ref{eq:finalrecursion}). The building blocks (i.e. integrals of the type $I[x]$) for calculation of 1 loop correction can be just guessed after some experience and are given by
 \be\label{cauchyintegrals}
 &&\intl_{-1}^1 dy \frac 1{y-x}=\log\[\frac{x-1}{x+1}\],\no\\
 &&\intl_{-1/b}^{1/b} dy \frac 1{y-x}=\log\[\frac{x-\frac 1b}{x+\frac 1b}\],\no\\
 &&\intl_{-1}^1 dy \frac 1{y-x}\sqrt{b^2-x^2}=\sqrt{b^2-x^2}\log[\frac{x\sqrt{b^2- 1}-\sqrt{b^2-x^2}}{x\sqrt{b^2- 1}+\sqrt{b^2-x^2}}],\no\\
 &&\intl_L dy \frac 1{y-x}\sqrt{b^2-\frac 1{x^2}}=\sqrt{b^2-\frac 1{x^2}}\log[\frac{x\sqrt{b^2-\frac 1{x^2}}-\sqrt{b^2-1}}{x\sqrt{b^2-\frac 1{x^2}}+\sqrt{b^2-1}}],\no\\
 &&\intl_L dy \frac 1{y-x}\frac{\sqrt{b^2-x^2}}{\sqrt{b^2-\frac 1{x^2}}}=\frac{\sqrt{b^2-x^2}}{\sqrt{b^2-\frac 1{x^2}}}\log[\frac{x\sqrt{b^2- \frac 1{x^2}}-\sqrt{b^2-x^2}}{x\sqrt{b^2- \frac 1{x^2}}+\sqrt{b^2-x^2}}],
\ee
where $L=[-1,-1/b]\cup[1/b,1]$.

One can check that the discontinuity of the r.h.s on the contour of integration coincides with the integrand in the l.h.s.

In view of the relation (\ref{f}), knowledge of the integrals (\ref{cauchyintegrals}) makes the calculation of $S_1$ straightforward.

\section{Integrals for the 2-loop correction}\label{sec:2loop}
Direct calculations shows that
\be
  -2\ell I_{s,1}&+&4\ell I_{s,2}\(1+\frac 1{2\ell^2}\)=\intl_{\g_+}\frac {dy}{2\pi}\frac{4y\(\frac{1-2b^2+y^2}{\sqrt{b^2-y^2}}+\frac{\frac 1{y^2}+1-2b^2}{\sqrt{b^2-\frac 1{y^2}}}+4\ell\)}{(y^2-1)^2}S_1[1/y]+\no\\
  &+&\intl_{\b_+}\frac{dy}{2\pi}\frac{4\sqrt{b^2-\frac 1{y^2}}(3+y^2(\ell^2-6)+y^4(4-\ell^2+2\ell^4))}{y^3(y^2-1)^2\ell^4}S_1[1/y],
\ee
\begin{footnotesize}
\be
  P&=&P_1+P_2,\no\\
  P_1&=&\frac 4{\ell^4}\intl_{\beta_+}\frac{dy}{2\pi}\frac{(y^2-1)(b^2-\frac 1{y^2})^{3/2}(-\frac d{dy}S1[1/y])-B_0}{1-b^2y^2},\no\\
   B_0&=&\frac {b^2}{2\ell^2\pi}(4b^4x\log[\frac{bx -1}{b x+1}]+(-2+2b-6b^2-2b^3+2(1-2b^2-2b^3+b^4)\log[b-1]-\no\\&&8b^2(1+b^2)\log[b]+(4b^2+4b^3+4b^4)\log[1+b]+(-1+4b^2+b^4)\log[1+b^2])),\no\\
  P_2&=&\frac{b\log 2b}{\ell^6\pi^2}(-2+2b-6b^2-2b^3-b^3\log[4]+2(1-2b^2-2b^3+b^4)\log[b-1]-\no\\&&2b^2(4+b+4b^2)\log[b]+(4b^2+4b^3+4b^4)\log[1+b]+(-1+4b^2+b^4)\log[1+b^2]).
\ee

\end{footnotesize}
  \bibliography{article3}        
 \bibliographystyle{utphys}
\end{document}